\documentclass[final,3p,times]{elsarticle}

\journal{Neural Networks}

\usepackage{amsmath,amssymb,amsfonts}
\usepackage{algorithmic}
\usepackage{algorithm}
\usepackage{graphicx}
\usepackage{booktabs}
\usepackage{tablefootnote}
\usepackage{threeparttable}
\usepackage{subfigure}
\usepackage{enumitem}
\usepackage{xcolor}
\usepackage{xspace}
\usepackage{multirow}
\usepackage{array}
\usepackage{tabularx}
\usepackage{adjustbox}
\begin{document}

\begin{frontmatter}

\title{Understanding of Task-specific and Subject-specific Components \\in Surface EMG}
\author[label1]{Yangyang Yuan}
\ead{yyyuan25@sjtu.edu.cn}
\author[label2]{Jionghui Liu}
\author[label3]{Xinyu Jiang}
\author[label1]{ChihHong Chou}
\author[label1]{Chenyun Dai}
\author[label4]{Jiahao Fan\corref{corresponding}}
\ead{jxf5683@psu.edu}

\cortext[corresponding]{Corresponding author.}

%% Author affiliation
\affiliation[label1]{organization={School of Biomedical Engineering, Shanghai Jiao Tong University},%Department and Organization
            % addressline={}, 
            city={Shanghai},
            postcode={200240}, 
            % state={},
            country={China}}
\affiliation[label2]{organization={Institute of Science and Technology for Brain-Inspired Intelligence, Fudan University},%Department and Organization
            % addressline={}, 
            city={Shanghai},
            postcode={200240}, 
            % state={},
            country={China}}
\affiliation[label3]{organization={School of Informatics, The University of Edinburgh},%Department and Organization
            % addressline={}, 
            city={Edinburgh},
            % postcode={200240}, 
            % state={},
            country={UK}}
\affiliation[label4]{organization={Department of Mechanical Engineering, the Pennsylvania State University},%Department and Organization
            % addressline={}, 
            city={PA},
            % postcode={200240}, 
            % state={},
            country={U.S.}}

%% Abstract
\begin{abstract}
%% Text of abstract
Surface electromyogram (sEMG) signals are widely used in human-machine interfaces for gesture recognition and user identification, but existing models often struggle with generalization across different individuals due to subject-specific neuromuscular characteristics. This study introduced a disentanglement model to separate task-specific and subject-specific components from sEMG signals, thus improving the generalization and interpretability of gesture recognition and user identification systems. Experimental results demonstrate that disentangled task-specific components significantly improve the accuracy of both gesture classification and user identification across different subjects and days, outperforming conventional methods in the same scenario. Further analysis of the extracted components reveals that task-specific components capture consistent activation patterns for the same gestures across individuals. In contrast, subject-specific components reflect unique neuromuscular characteristics that can be used for user identification. Notably, subject-specific components show reduced similarity compared to task-specific components in inter-day scenarios, contributing to more accuracy decrease in user identification than in gesture recognition. These findings suggest that the disentanglement approach not only boosts classification performance but also provides deeper insights into the physiological mechanisms underlying sEMG signals. The model’s ability to isolate and interpret different neuromuscular components holds promise for enhancing the robustness of sEMG-based applications in real-world settings, such as rehabilitation and user authentication.
Our code is available at: https://github.com/Open-EXG/HandDisentanglement.
\end{abstract}

% %%Graphical abstract
% \begin{graphicalabstract}
% %\includegraphics{grabs}
% \end{graphicalabstract}

%%Research highlights
% \begin{highlights}
% \item A neural network inspired by clinical priors and human brains for EEG-MI decoding.
% \item Lightweight attention for weighted EEG spatial-spectral information.
% \item Efficient module for multi-scale long-term temporal information extraction.
% \item Dual prototype learning for enhancing model generalization ability.
% \end{highlights}

%% Keywords

\end{frontmatter}

\section{Introduction}
Surface electromyogram (sEMG) is one of the common neural signals in the practical applications of human-machine interface \cite{zheng2022surface}, since the neural signals acquired on the skin surface can reflect human's movement intention. Accordingly, the intelligent machines embedded with advanced models recognize the movement intention from their sEMG to realize the intuitive interactions between human and machines \cite{xiong2021deep}. In all the sEMG-based human-machine interactions, hand gesture \cite{hu2018novel, su2021hand} is the most widely used interface in the related applications (e.g., rehabilitation \cite{guo2019method, yang2022task}, neuroprosthesis \cite{yeon2021acquisition}, biomechanics \cite{disselhorst2020surface}, etc.). The sEMG signals generated from forearm muscles when users perform different gestures are used as the command inputs of human-machine systems. Therefore, accurately interpreting sEMG for different hand gestures is the key to the effectiveness of such an interface. 

For decades, an enormous amount of research \cite{khezri2007real, zhuojun2015semg, bittibssi2021semg, dai2019extracting} has put great efforts into establishing the pattern recognition models between sEMG and movement intention. However, most of the studies aim to develop a subject-specific model. The hand gesture recognition accuracy may substantially decrease when a well-established model is directly applied to a new subject \cite{meng2022user}. The poor generalization ability lowers the practical value of the technique, since any new subject requires a long and cumbersome adaptation procedure for model calibrations \cite{zhai2017self, nasri2020semg, guo2024semg} (e.g., pre-trained data acquisition, model training, etc). Fortunately, many recent studies have validated the feasibility of developing a subject-generalized model, since the anatomical structure of muscles and nerves in the human forearm is roughly identical across individuals \cite{kanoga2020subject, suh2023tasked}. Based on the hypothesis that the amplitude range of sEMG signals is roughly consistent across different subjects, Lin et al. \cite{Lin2020Normal} proposed a concise method using max-min normalization, enhancing the model's performance across users by over 20\% compared to the baseline. In the field of advanced deep learning algorithms, transfer learning has been used in model generalization as well. Chen et al. \cite{Chen2021Hand} applied an effective transfer learning strategy in a traditional convolutional neural network (CNN)-based model, improving the recognition accuracy by 10\%--38\% on the target dataset from new users.

Although the recognition accuracy of the subject-generalized model has been significantly improved, it still cannot achieve the equivalent accuracy level of the subject-specific model \cite{jiang2016development}. The main reason is that the subject-specific component exists in sEMG due to the differences in neuromuscular structures and force generation habits across individuals. Many recent studies \cite{jiang2020neuromuscular} even found that the subject-specific component in sEMG signals can be developed to be a new biometric modality for user identification. Corresponding to the subject-specific component, establishing a subject-generalized model can be accordingly regarded as extracting the task-specific component in sEMG across individuals. Therefore, one solution is to establish a model that disentangles task-specific and subject-specific components in sEMG signals.

The idea for developing this disentanglement model is inspired by style transfer learning techniques \cite{gupta2019image} established in the field of computer vision. In the image recognition task, the same content or object but with different painting styles in images still needs to be identically recognized \cite{deng2022stytr2, kwon2022clipstyler}. Accordingly, a framework of the proposed model is built based on a two-encoder and one-decoder architecture \cite{aberman2019learning, gu2020cross}. The two encoders of the original model can separate the object in images into "content" representation and "style" representation.
In a similar manner, the task-specific and subject-specific components in sEMG can be viewed as "content" and "style," respectively. Analogically, in our work, the two encoders disentangle the sEMG into task-specific and subject-specific components. In addition, the decoder reconstructs the model input using the two learned latent representations. Our previous work \cite{fan2024surface} firstly proposed the concept of sEMG feature disentanglement and extracted task-specific and subject-specific representation. However, in this research, we only utilized the two extracted representations for improving the generalization of gesture recognition and user identification model, in absence of physiological interpretation for the disentangled components. 

In this work, based on the above findings, we established a disentanglement model to extract both the task-specific and subject-specific components. The work was validated on sEMG data of 20 \textcolor{black}{subjects} collected from two different days for 11 1-degree-of-freedom (1-DoF) gesture recognition.  Furthermore, the disentangled feature patterns of two components were systematically quantified and delved into their physiological interpretation. This work represents the first evidence that neural network models can interpretatively analyze neuromuscular information.

\section{Related Works}

\textcolor{black}{In EMG-based human-machine interaction systems, extracting generalized EMG features applicable across various scenarios remains challenging. Variations in EMG signals arising from different gestures, subjects, and acquisition conditions can affect the generalizability of models across scenarios. As a result, task-specific calibration steps are often indispensable. With labeled data available for additional training and calibration, in gesture recognition tasks, most researchers have chosen to fine-tune certain modules (typically the classification layer) of the model according to new sEMG features, thus enhancing its classification capability for different gestures from new users. For example, Kim et al. \cite{kim2019subject} trained a convolutional neural network (CNN) and fine-tuned the model for the new user, improving the classification accuracy by 23.2\%. In a follow-up study conducted by Wang \cite{wang2023deep}, attention modules were integrated into the convolutional network to focus the fine-tuning procedure on more important layers instead of specific ones. In user identification tasks, Jiang et al. used the combination of gestures and “authentication + identification” to improve the recognition accuracy in inter-day scenarios \cite{jiang2021enhancing,jiang2022measuring}, which also relies on labeled samples.}

\textcolor{black}{In cases where labeled data is unavailable, most current research focuses on obtaining sEMG features that minimize inter-user differences to improve gesture recognition accuracy across different users. One of the most representative approach in this scenario is domain adaptation. Zhang et al. \cite{zhang2022domain} utilized the domain distance in a kernel space as an indicator to screen out reliable instantaneous samples for updating the feature extractor, thus improving the alignment of feature representations of myoelectric patterns across users. To address the sEMG distribution variations between disparate training users, Zhang et. al. proposed a multi-source synchronize domain adaptation framework to align each source user and the new user in individual feature spaces, further improving the model's inter-user performance \cite{zhang2023multi}. However, although supporting a small amount of unlabeled data, unsupervised transfer learning algorithms still require additional data and training for calibration, and is difficult to be applied in identification tasks.}

\textcolor{black}{Therefore, we urged for more universal feature extractor to obtain EMG representations effective in multiple scenarios. Holobar et al. \cite{holobar2014accurate} utilized neural decomposition to isolate the discharges of individual motor units from EMG signals, aiming to obtain universal microscopic EMG representations for prosthetic hand control. For macroscopic EMG features, Han et al. \cite{han2020disentangled,han2021universal} employed two adversarial networks to respectively minimize or maximize the predictability of subject identity, guiding the feature extractor to suppress or enhance subject-specific information. This enabled a clearer separation between task-specific and subject-specific feature components. The extracted task-specific components improved the inter-user accuracy of gesture recognition by 11.6\%. Inspired by these works, along with the related works on computer vision \cite{aberman2019learning} and gait recognition domain \cite{gu2020cross}, a disentanglement network based on multi-encoder and single-decoder framework was proposed to extract the fundamental subject-specific and task-specific components from EMG signals.}

\begin{figure}
    \vspace{0.6cm}
    \centering
    \includegraphics[width=0.7\linewidth]{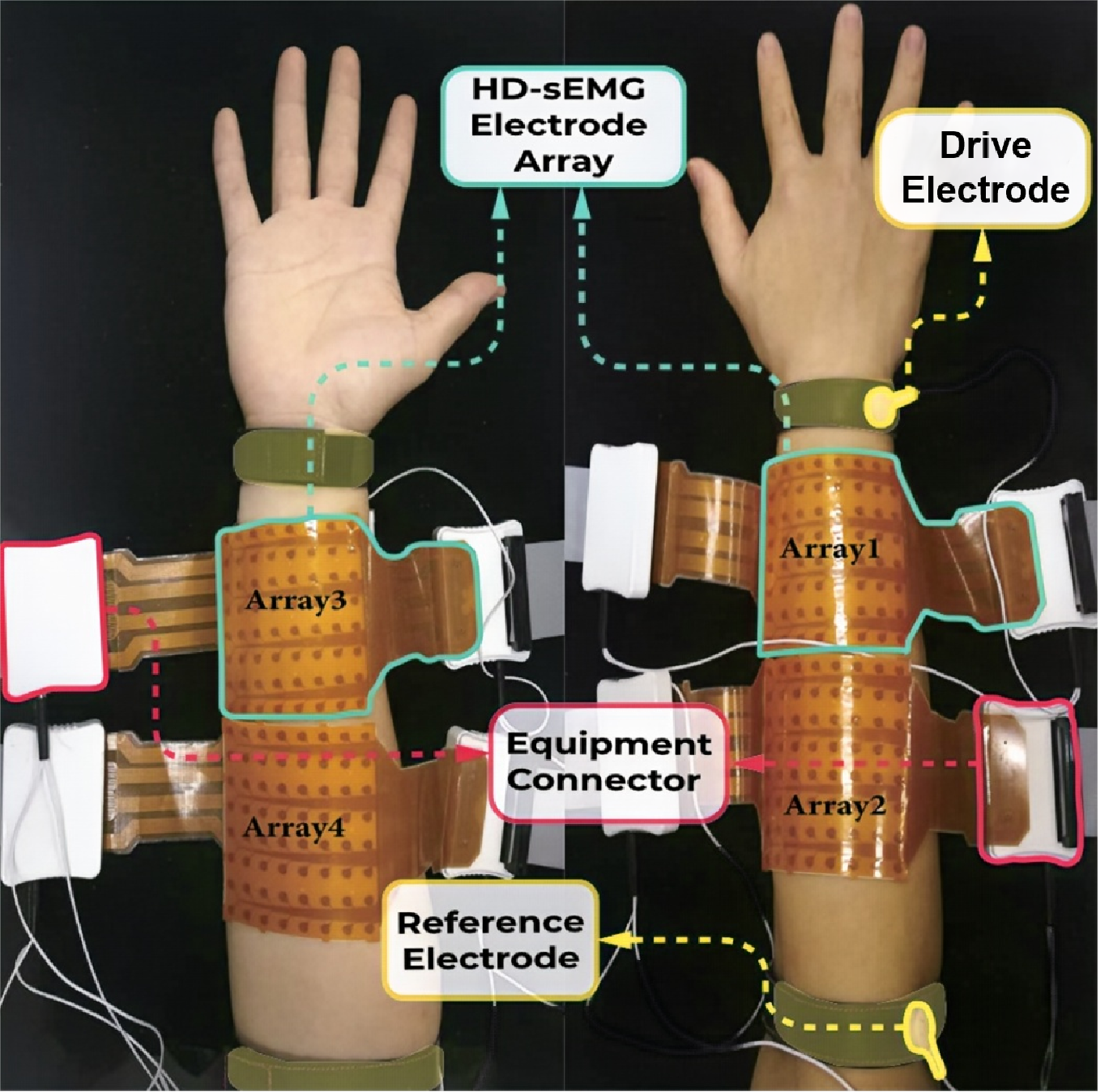}
    \caption{Electrode setup in the experiment.}
    \label{fig:electrode_setup}
\end{figure}

\section{Materials}
We validated the proposed method on the open source dataset from our previous work \cite{jiang2021open}, which can be accessed at the website \footnote{(https://doi.org/10.13026/ym7v-bh53)}. In this study, we only used the pattern recognition subset of the dataset for analysis. In the following, we \textcolor{black}{provided} a brief introduction about the subject information and data acquisition.

\subsection{Subjects}
The experiment invited 20 subjects, with 8 females and 12 males aged between 22 and 34 years. All participants were thoroughly informed about the experimental procedures and \textcolor{black}{were} provided written informed consent. The study was conducted under the supervision and approval of the ethics committee of Fudan University (approval number: BE2035).

\subsection{Data Acquisition}

Four gelled electrode arrays were mounted on the forearm of the subject's dominant hand for sEMG data collection, with two 8\texttimes8 arrays (16\texttimes8 electrodes) on the flexor and the other two on the extensor respectively. Each electrode is elliptical, with the major axis 5 mm, the minor axis 2.8 mm and the inter-electrode distance 10 mm. \textcolor{black}{We placed the reference electrode on the olecranon and the drive electrode around the wrist.} To ensure the maximum coverage of the forearm muscles, we aligned the center of each 16\texttimes8 electrode array with the center of the flexor or extensor. Figure \ref{fig:electrode_setup} illustrates the details of the electrode setup. For better signal quality, subjects were required to clean the skin with an alcohol wipe before placing the electrodes to reduce the contact impedance.

\begin{figure}
    \vspace{0.6cm}
    \centering
    \includegraphics[width=0.9\linewidth]{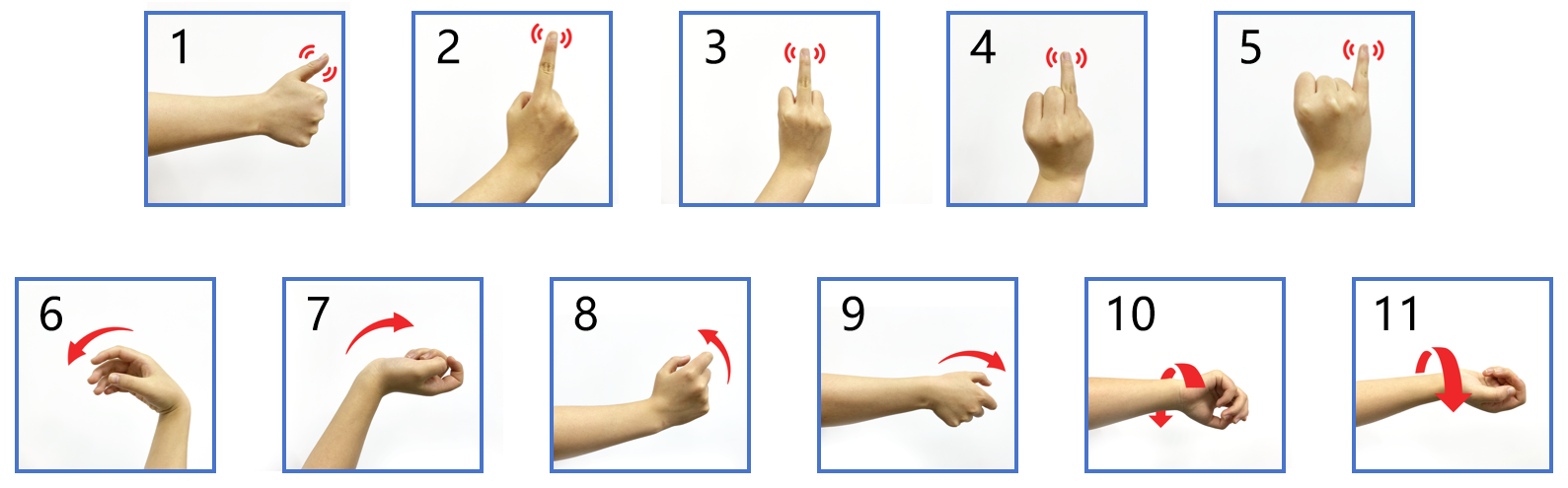}
    \caption{Gestures involved in the validation.}
    \label{fig:gestures}
\end{figure}

During the experiment, the subjects were required to sit in a \textcolor{black}{quiet} room and perform specified gestures with their dominant hand at their comfortable speed (approximately 1-second), following the instructions on the screen in front of them. The subjects were required to perform 11 common gestures shown in Figure \ref{fig:gestures} with 6 repetitions. We \textcolor{black}{provide} a 2-second inter-trial break and a 5-second inter-gesture break to avoid cumulative muscle fatigue. 

The selected 11 gestures for further analysis were (1) thumb extension, (2) index extension, (3) middle extension, (4) ring extension, (5) little extension, (6) wrist flexion, (7) wrist extension, (8) wrist radial, (9) wrist ulnar, (10) wrist pronation, and (11) wrist supination. We prefer these 11 gestures because most of them are 1-degree-of-freedom (1-DoF) motions. They can present more significant and intuitive feature patterns of task-specific components after disentangled, since the interference of other DoFs is eliminated. Additionally, the 11 gestures were widely used in daily life and included all the finger and wrist motion freedoms. \textcolor{black}{Further more, for multi-DoF gestures, the complexity and variety of types make it difficult to achieve exhaustive data collection and research. However, a previous study\cite{yuan2024training} has shown that multi-DoF hand movements can be obtained by linearly combining single-DoF movements. Therefore, using single-DoF gestures can adapt to multi-DoF scenarios flexibly by generating multi-DoF sEMG features through the superposition of single-DoF gestures, and also provide a clearer understanding of muscle activation patterns for task-specific and subject-specific components by reducing interference between muscles.}

The above data collection procedures were performed for all subjects on two different days, with a time interval of 3 to 25 days. Overall, we acquired a total of 1320 (11 gestures\texttimes 6 repetitions\texttimes 20 subjects) trials for analysis in this study.

\section{Methods}
\subsection{Signal Preprocessing}
The acquired raw sEMG data were first filtered by a 10--500 Hz band-pass filter and a series of power-line interference removal notch filters at 50 Hz and its harmonics up to 400 Hz. Then, the continuous filtered sEMG data were segmented into motion samples and rest samples according to the trigger recorded during the experiment. The duration of each sample involved in this study is 1 second. 

\subsection{Feature Extraction}
For the model training, the four sEMG features commonly used in many previous studies \cite{dai2016comparison,jiang2021open} were selected to extract sEMG representations for each sample, namely root mean square (RMS), wave length (WL), zero crossing (ZC) and slope sign change (SSC). Since each 16\texttimes8 electrode arrays (128 channels) were used to collect sEMG data for either extensor or flexor muscle, \textcolor{black}{the model inputs were reshaped into two 128 feature vectors}.

\textcolor{black}{In a previous study \cite{yuan2024exploring}, we compared the effects of three different categories of EMG measures on the disentanglement effect. Among the individual measures, the frequency-domain measure STFT exhibited the best performance, followed by time-domain measures (RMS, WL, SSC, ZC), with waveform information showing the least effectiveness. However, in terms of physiological interpretability, the heatmap distribution of time-domain measures aligns more closely with actual muscle activation patterns than that of frequency-domain measures.} For the clarity of result presentation and visualization, we only used the RMS feature. \textcolor{black}{RMS is a representative energy descriptor to provide a direct observation and physiological interpretation on the task-specific and subject-specific patterns, which can be described as:}
\begin{equation}
   \textcolor{black}{RMS=\sqrt{\frac{1}{N}\sum_{i=1}^{N}x_{i}^{2}}} 
\end{equation}
\textcolor{black}{where $N$, $x_i$ respectively denote the total number of the sample point in a window and the signal value of the i-th sample point in a single channel.} The RMS was extracted from each channel of the two 16$\times$8 electrode arrays. Then, the RMS heat maps of the two 16$\times$8 electrode arrays correspond to the muscle activation of the extensor and flexor, respectively.

To address the issue related to corrupted channels, we implemented an outlier recovery for each feature along the spatial axis. Specifically, one certain channel was considered as an outlier if its RMS deviated by more than three standard deviations from the mean value of all channels. Such outliers were subsequently replaced with the average value of their adjacent channels. The recovery procedure of corrupted channels was performed independently for the signal segments of each task. After the data pre-processing and feature extraction steps were completed, we attempted to search for a generic latent subspace where the sEMG features of each hand gesture across all the subjects were maximally differentiated. 

\subsection{Feature Disentanglement}

We decomposed the sEMG feature into two orthogonal subspaces--task-specific (subject-invariant) and subject-specific (task-invariant) subspaces. Let $\mathcal{P}$ and $\mathcal{S}$ denote the set of different gestures and subjects, respectively. Let $x_{i,j}$ represent a sample that is characterized by two attributes: the corresponding gesture task $i\in \mathcal{P}$ and the identity of the subject $j\in \mathcal{S}$. \textcolor{black}{The fundamental concept of our methodology is to establish two orthogonal latent spaces for the aforementioned two attributes}: a task-specific latent space where samples representing similar hand gestures across different subjects are tightly clustered, and a subject-specific latent space where the discrepancies within the same subject are minimal compared to variations between different subjects. This approach enables us to derive two distinct representations from the original feature space that are disentangled from each other.

\begin{figure*}
    \centering
    \includegraphics[width=0.7\linewidth]{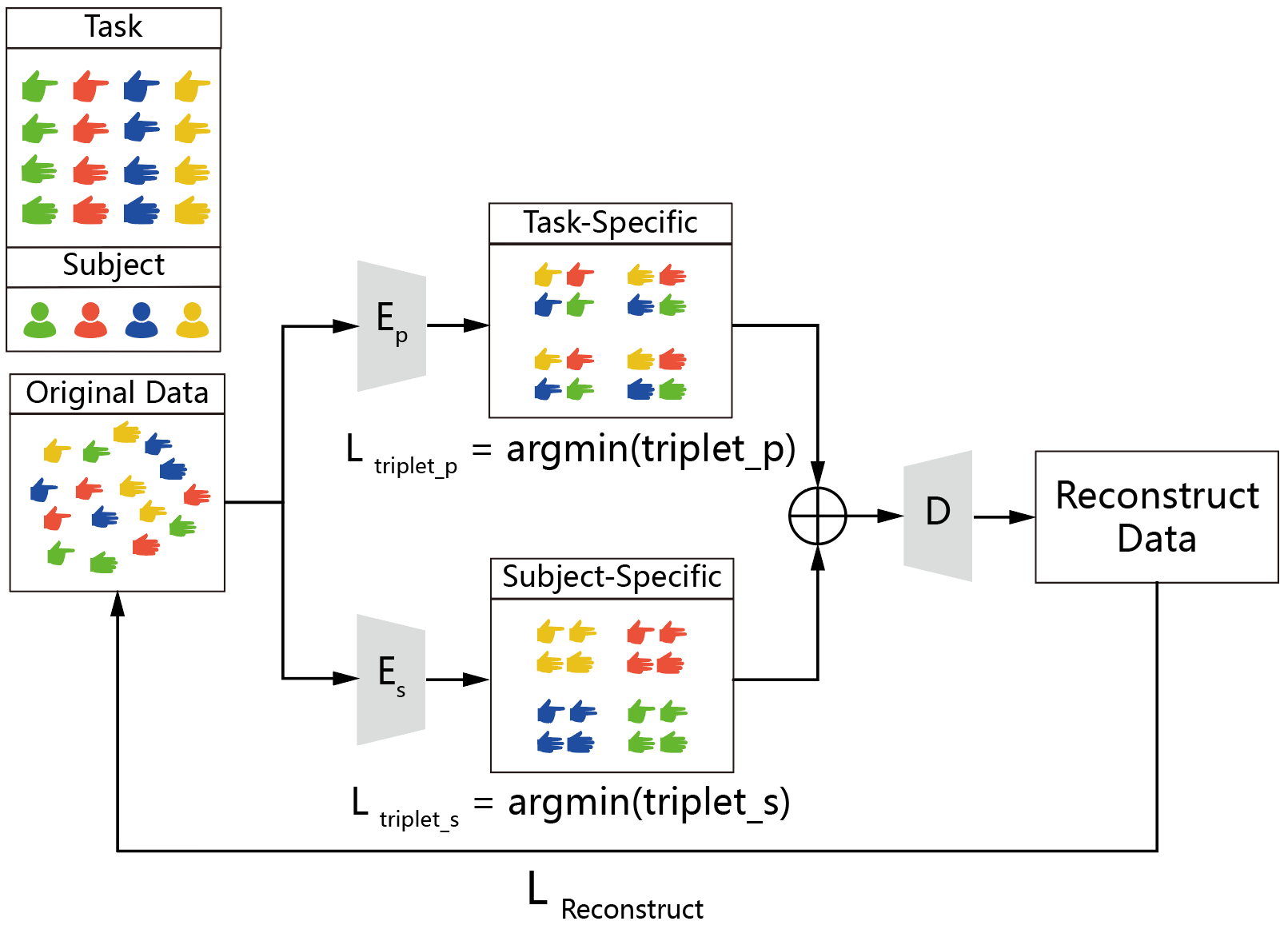}
    \caption{The framework of the mode, composed of two encoders and one decoder. $E_{p}$ and $E_{s}$ denote the task-specific encoder and the subject-specific encoder respectively.}
    \label{fig:ModelStructure}
\end{figure*}

As illustrated in Figure \ref{fig:ModelStructure}, our method was implemented by an autoencoder-like neural network, which consists of two encoders and a single decoder. The encoder $E_{p}$ was used to project the original input $x_{i,j}$ into the pattern-related latent space, yielding a task-specific representation $E_p({x_{i,j}})$, whereas \textcolor{black}{the encoder $(E_{s})$} was trained to extract the subject-specific representation $E_s({x_{i,j}})$. The output of these two encoders was concatenated and fed into a decoder $D$ to reconstruct the input from the latent space representation.

To train the network, we used a specifically designed loss function which consists of three parts. First, a reconstruction loss is used to ensure the decoder can reconstruct the original input from the two latent representations:
\begin{equation}
    \mathcal{L}_{recon}=\mathbb{E}[\Vert D(E_{p}(x_{i,j}),E_{s}(x_{i,j}))-x_{i,j}\Vert]
\end{equation}

The second part is the triplet loss, which adds a constraint to the learned latent representations of the two encoders to promote disentanglement. The key idea was that the distance between two samples that share the same attributes should have a much smaller distance than two samples with different attributes. For example, on the task-specific latent space, we hope that the representation of samples with the same hand gesture could be similar, regardless of the subject. To achieve this, we used a triplet loss consisting of $\mathcal{L}_{trip\_p}$ and $\mathcal{L}_{trip\_s}$:
\begin{equation}
\begin{split}
    \mathcal{L}_{trip\_p}&=\mathbb{E}[\Vert E_{p}(x_{i,j})-E_{p}(x_{l,j})\Vert-\Vert E_{p}(x_{i,j}) \\
    &-E_{p}(\textcolor{black}{x_{l,k}})\Vert+\alpha ]_{+}
\end{split}
\end{equation}
\textcolor{black}{where} $l\neq i, k\neq j$. \textcolor{black}{$x_{l,j}$} represents samples that share the same gesture with $x_{i,j}$ but was from a different subject, whereas \textcolor{black}{$x_{l,k}$} represents samples that have a different gesture with $x_{i,j}$, and $\alpha$ denotes the margin. $\textcolor{black}{\mathbb{E}}[x]_{+}=max\{x,0\}$ selected the larger value between $x$ and zero. $\Vert E_{p}(x_{i,j})-E_{p}(x_{l,j})\Vert$ measures the distance between the same gestures from different subjects, while $\Vert E_{p}(x_{i,j})-E_{p}(x_{l,k})\Vert$ measures the distance between different gestures from different subjects. During training, the model aims to minimize the former while maximizing the latter, ensuring that the same gestures are clustered together while different gestures are separated. $\alpha$ is a non-negative value that enforces the distance between different gestures to be larger than that of the same gestures by at least this value. The same triplet loss $\mathcal{L}_{trip\_s}$ is also applied to the subject latent space. The total triplet loss can be described as:
\begin{equation}
    \textcolor{black}{\mathcal{L}_{trip}=\mathcal{L}_{trip\_p}+\mathcal{L}_{trip\_s}}
\end{equation}

The last part is the cross-reconstruction loss, which is formulated as :
\begin{equation}
    \begin{split}
        \mathcal{L}_{cross}&=\mathbb{E}[\Vert D(E_{s}(\textcolor{black}{x_{i,k}}),E_{p}(\textcolor{black}{x_{l,j}}))-(x_{i,j})\Vert]
    \end{split}
\end{equation}
\textcolor{black}{where} $l\neq i, k\neq j$. The cross-reconstruction loss encourages $E_p$ and $E_s$ could generate latent representations that can be mapped back into the ground truth samples that share similar attributes when the two representations are cross-combined. 

Summing the above three terms, we obtain the total loss as:
\begin{equation}
\mathcal{L}=\mathcal{L}_{recon}+\lambda_{1}\mathcal{L}_{cross}+\lambda_{2}\mathcal{L}_{trip}
\end{equation}
\textcolor{black}{where} $\lambda_{1}$ and $\lambda_{2}$ are the balance weights. \textcolor{black}{Different weights were tried in the \textcolor{black}{preliminary} experiment, the results showed that their values were not sensitive on recognition accuracy ($<$ 1\% difference). To ensure that $\mathcal{L}_{trip}$ and $\mathcal{L}_{cross}$ have comparable values, we set $\lambda_{1}=1, \lambda_{2}=0.5$ in this study.}

\begin{table}
\centering
\renewcommand{\arraystretch}{1.5}%
\caption{\protect\label{tab:network_architecture}The table summarizes the parameters of the network used in this study. Conv, IN, LRLU, UpS, RP, and DO represent Convolution, Instance Normalization, Leaky ReLU, Upsampling, Reflection Padding, and Dropout layers, respectively. The kernel size, stride, and padding of the Convolution layer are denoted by $k$, $s$, and $p$, respectively. \textcolor{black}{$d$ denotes the dimension of the sEMG features input into the encoder and output from the decoder.} In/Out refers to the number of input and output channels for each module. The slope of the Leaky ReLU activation function and the Dropout probability are both set to 0.2.}
\begin{tabular}{ccc}
\hline
\textbf{Module}     & Encoder       & Decoder                                                        \\ \hline
\textbf{}           & Conv+IN+LRLU  & \begin{tabular}[c]{@{}c@{}}UpS+RP+Conv\\ +DO+LRLU\end{tabular} \\
\textbf{Layers}     & Conv+IN+LRLU  & \begin{tabular}[c]{@{}c@{}}UpS+RP+Conv\\ +DO+LRLU\end{tabular} \\
\textbf{}           & Conv+IN+LRLU  & \begin{tabular}[c]{@{}c@{}}UpS+RP+Conv\\ +DO+LRLU\end{tabular} \\ \hline
\textbf{Parameters} & $k$=3, $s$=2, $p$=1 & $k$=3, $s$=1, $p$=1                                                  \\ \hline
\textbf{In/Out}     & $d$/128         & 256/$d$                                                          \\ \hline
\end{tabular}
\end{table}

The details of the utilized network are described in Table \ref{tab:network_architecture}. In our implementation, all of the components are based on 2-D convolution layers. For encoder $E_p$ and $E_s$, the input underwent a downsampling process by a convolution stride of 2, \textcolor{black}{where the kernel size (k), step size (s) and padding size (p) were determined according to the input and output size of data (with the input size of $16\times16$ and output size of $2\times2$) and the muscle activation area that the kernel size can cover.} Instance normalization was applied to eliminate the influence of the various intensities of the features from different subjects, as we care more about the structural information of the EMG feature. For decoder $D$, a linear interpolation is used to upsample the latent representation to map the original input. \textcolor{black}{Among the hyperparameters such as the slope of LeakyReLU and the dropout probability in the network, only the slope of LeakyReLU had a significant impact on model performance, which decreases the classification accuracy as it increases when exceeding 0.2.}

During training, data augmentation \cite{jiang2022optimization} was performed by randomly translating and rotating the original features from each $8 \times 8$ electrode array. \textcolor{black}{These translation and rotation strategies were selected because electrode shift between experiment sessions and different subjects is inevitable during data acquisition. The translation distance is randomly sampled from a uniform distribution ranging from -15 mm to 15 mm (positive values indicate rightward and upward shifts), while the rotation angle is randomly sampled from a uniform distribution ranging from -15° to 15° (positive values indicate clockwise rotation). Each original sample undergoes one translation and one rotation augmentation, resulting in a final sample size three times that of the original dataset.} The network weights were optimized using the Adam optimizer \cite{kingma2014adam} with an initial learning rate of $0.002$, which was gradually reduced by a factor of 0.5 every 200 iterations. Training was conducted for a maximum of 500 iterations with a batch size of 2000. A dropout probability of 0.8 was applied to the decoder layers to prevent overfitting.

\subsection{Experimental Protocols}
\subsubsection{Training-testing Validations}
In this study, two training-testing validations were performed: 1) intra-day validation: A $k$-fold cross-validation was implemented to ensure even distribution across all classes. In each fold, two-thirds of the samples from each gesture of all subjects were used for training, and the rest for testing. Two independent runs were conducted, each using data from Day 1 and Day 2, respectively. 2) inter-day validation: In the inter-day validation, samples from all the subjects acquired on Day 1 were used to train the models. The models were then tested on data collected on Day 2.  

\subsubsection{Classification Accuracy}
First, our method was evaluated on classification performance of disentanglement model. We assessed the performance of task- and subject-specific components derived from sEMG in their respective downstream applications: task-specific components in gesture recognition and subject-specific components in identity recognition. Since the disentanglement model is essentially a feature extraction model, we used the simplest classifier, k-Nearest Neighbor (KNN) for gesture or identity recognition after feature extraction. This demonstrates that the model proposed in this study is capable of extracting the two components on its own, without relying on the classification ability of a powerful classifier. In addition, the classification performance of proposed disentanglement model was compared with three baseline models, namely original feature, principal component analysis (PCA) and standard autoencoder (AE). The original feature referred to directly using the extracted features from the sEMG data without any additional signal processing procedures. The PCA and AE \textcolor{black}{are} two commonly used models which can project the original features into task- and subject-specific components.

\subsubsection{Visual Inspection}
Second, to elucidate the potential physiological significance of these components, we visually displayed the heat maps of disentangled task- and subject-specific components. For each component, we showed two 16$\times$8 heat maps, which had the same topological structure as the electrode channels shown in Figure \ref{fig:electrode_setup}. This mapping enabled to link the disentangled components to the underlying neuromuscular patterns shared across individuals and identify potential biomarkers within each subject. We exhibited the disentangled task- and subject-specific components for all the 11 gestures of all the 20 subjects. Then, we averaged the heat map of task-specific components across subjects, and averaged that of subject-specific components across gestures. In addition, the visual inspection of proposed disentanglement model was compared with the RMS heat map of 2-dimensional EMG activation.

\begin{table*}[t!]
\centering
\renewcommand{\arraystretch}{1.5}%
\caption{The gesture recognition (11 gestures) performance under intra-day and inter-day validation protocol. The values in bold indicate the best results for each validation protocol.}
\label{tab:pattern}
\setlength{\tabcolsep}{1.2mm}{
\begin{tabular}{ccccccccccccc}
\hline
\multirow{2}{*}{Methods} & \multicolumn{4}{c}{day 1}                                         & \multicolumn{4}{c}{day 2}                                         & \multicolumn{4}{c}{Inter-day}                                     \\ \cline{2-13} 
                         & Acc.           & F1             & Pre            & Rec            & Acc.           & F1             & Pre            & Rec            & Acc.           & F1             & Pre            & Rec            \\ \hline
Orig. space              & 92.43          & 92.46          & 92.6           & 92.41          & 95.43          & 95.40          & 95.45          & 95.43          & 74.42          & 74.46          & 75.33          & 74.3           \\
PCA                      & 92.20          & 92.23          & 92.39          & 92.17          & 95.43          & 95.40          & 95.44          & 95.43          & 74.88          & 74.84          & 75.36          & 74.77          \\
AE                       & 95.75          & 95.76          & 95.79          & 95.76          & 97.29          & 97.28          & 97.3           & 97.28          & 77.98          & 77.69          & 77.93          & 77.87          \\
Proposed                 & \textbf{98.53} & \textbf{98.52} & \textbf{98.54} & \textbf{98.52} & \textbf{97.76} & \textbf{97.76} & \textbf{97.78} & \textbf{97.76} & \textbf{91.47} & \textbf{91.49} & \textbf{91.72} & \textbf{91.46} \\ \hline
\end{tabular}}
\end{table*}

\begin{table*}[t!]
\centering
\renewcommand{\arraystretch}{1.5}%
\caption{The identification recognition (20 identities) performance under intra-day and inter-day validation protocol. The values in bold indicate the best results for each validation protocol.}
\label{tab:identity}
\setlength{\tabcolsep}{1.2mm}{
\begin{tabular}{ccccccccccccc}
\hline
\multicolumn{1}{r}{\multirow{2}{*}{Methods}} & \multicolumn{4}{c}{day 1}                                      & \multicolumn{4}{c}{day 2}                                         & \multicolumn{4}{c}{Inter-day}                                     \\ \cline{2-13} 
\multicolumn{1}{r}{}                         & Acc.          & F1            & Pre           & Rec            & Acc.           & F1             & Pre            & Rec            & Acc.           & F1             & Pre            & Rec            \\ \hline
Orig. space                                  & 91.89         & 91.9          & 92.25         & 91.9           & 93.80          & 93.84          & 94.06          & 93.83          & 50.7           & 46.51          & 52.35          & 50.71          \\
PCA                                          & 91.74         & 91.74         & 92.1          & 91.73          & 93.57          & 93.61          & 93.81          & 93.6           & 50.47          & 46.81          & 51.02          & 50.49          \\
AE                                           & 94.98         & 94.97         & 95.09         & 94.96          & 96.43          & 96.46          & 96.53          & 96.45          & 52.79          & 50.71          & 52.27          & 52.92          \\
Proposed                                     & \textbf{99.3} & \textbf{99.3} & \textbf{99.3} & \textbf{99.31} & \textbf{98.84} & \textbf{98.85} & \textbf{98.87} & \textbf{99.84} & \textbf{64.65} & \textbf{62.48} & \textbf{67.90} & \textbf{64.76} \\ \hline
\end{tabular}}
\end{table*}

\subsubsection{Quantitative Evaluation}
Third, we selected several quantitative evaluation metrics to assess the two disentangled components. The silhouette (SIL) score \cite{rousseeuw1987silhouettes} was used to evaluate classification performance, which measures how similar a sample is to its own class compared to other classes. The average SIL value of all testing samples directly reflects whether different gestures using task-specific components, or different identities using subject-specific components, can be well distinguished. Moreover, we quantified the heat maps of the two components using the centroid coordinates ($C_x$, $C_y$), where $C_x$ represents the medial-lateral direction of the forearm, and $C_y$ represents the distal-proximal direction of the forearm.

\subsection{Statistical Analysis}
This study conducted statistical analysis on the accuracy of gesture recognition and identity identification. Since the data did not meet the assumption of normal distribution, a non-parametric chi-square test was used. A significance level was set at $p<0.05$.

\section{Results}

\subsection{Model Performance in User Identification and Gesture Recognition}

Table \ref{tab:pattern} details the gesture recognition performance with the extracted task-specific components. Our methods outperformed all other methods across all validation protocols, with the most notable performance improvement observed in the inter-day validation, where accuracy improved from 74.42\% (using original features) to 91.47\%. The significance of these accuracy improvements was confirmed by Friedman Tests (Day 1: $\chi^2 (3)=31.63, p<0.001$; Day 2: $\chi^2 (3)=7.63, p<0.05$, and Inter-day:  $\chi^2 (3)=26.39, p<0.001$). \textcolor{black}{Although the proposed method did not show a significant advantage over AE on Day 2 ($p>0.05$), \textit{post-hoc} analysis revealed that the proposed method significantly improved the classification accuracy in all other cases.} This indicates the efficacy of the task-specific components for the downstream pattern recognition tasks.

Table \ref{tab:identity} offers a comprehensive comparison in identity recognition results with the extracted subject-specific components. Notably, the intra-day identification accuracies are high for all approaches ($>$90\%), yet they experience a significant drop under inter-day validation. Our method consistently achieved the highest accuracy across all validation protocols. Statistical analysis revealed significant differences in accuracy between methods (Day 1: $\chi^2(3) = 36.05, p < 0.001$; Day 2: $\chi^2(3) = 35.38, p < 0.001$; Inter-day: $\chi^2(3) = 13.38, p < 0.01$). Further \textit{post-hoc} analyses confirmed that our method consistently outperformed all other methods, indicating that the subject-specific components maintained more robust representations linked to subject characteristics.

\subsection{Latent Feature Visualization}
We evaluated the utility of the disentangled task- and subject-specific components in two downstream applications: gesture recognition and user identification, under both intra-day and inter-day validation protocols.

\begin{figure}
    \centering
    \includegraphics[width=1.0\linewidth]{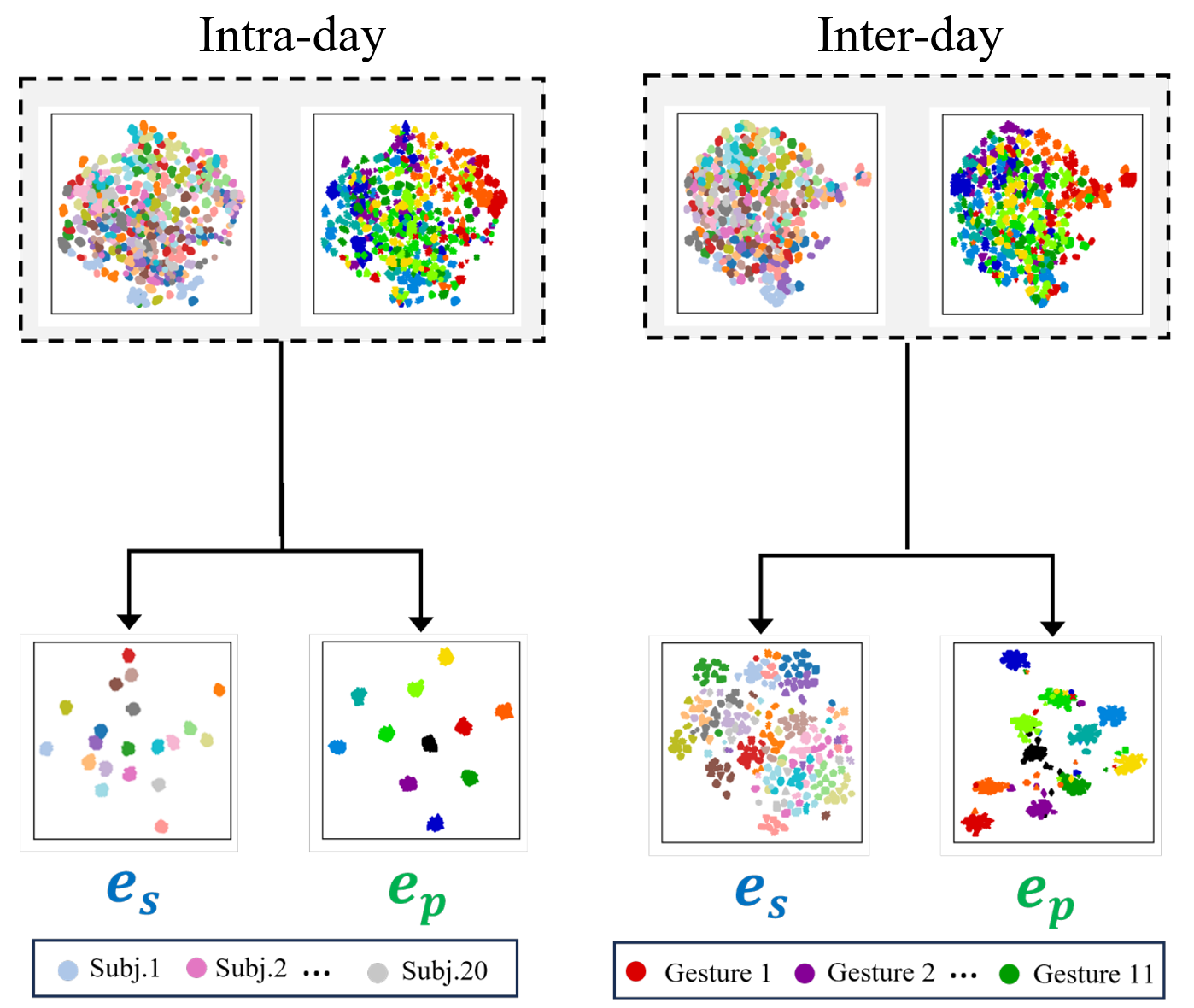}
    \caption{The visualization of original features (top row, with two figures representing features labeled by identities and gestures, respectively) and the disentangled latent vectors $e_p$ and $e_s$ (bottom row, with $e_s$ labeled by identities and $e_p$ labeled by hand gestures) using t-SNE. }
    \label{fig:disentangle_tsne}
\end{figure}

\begin{figure}
    \centering
    \includegraphics[width=1\linewidth]{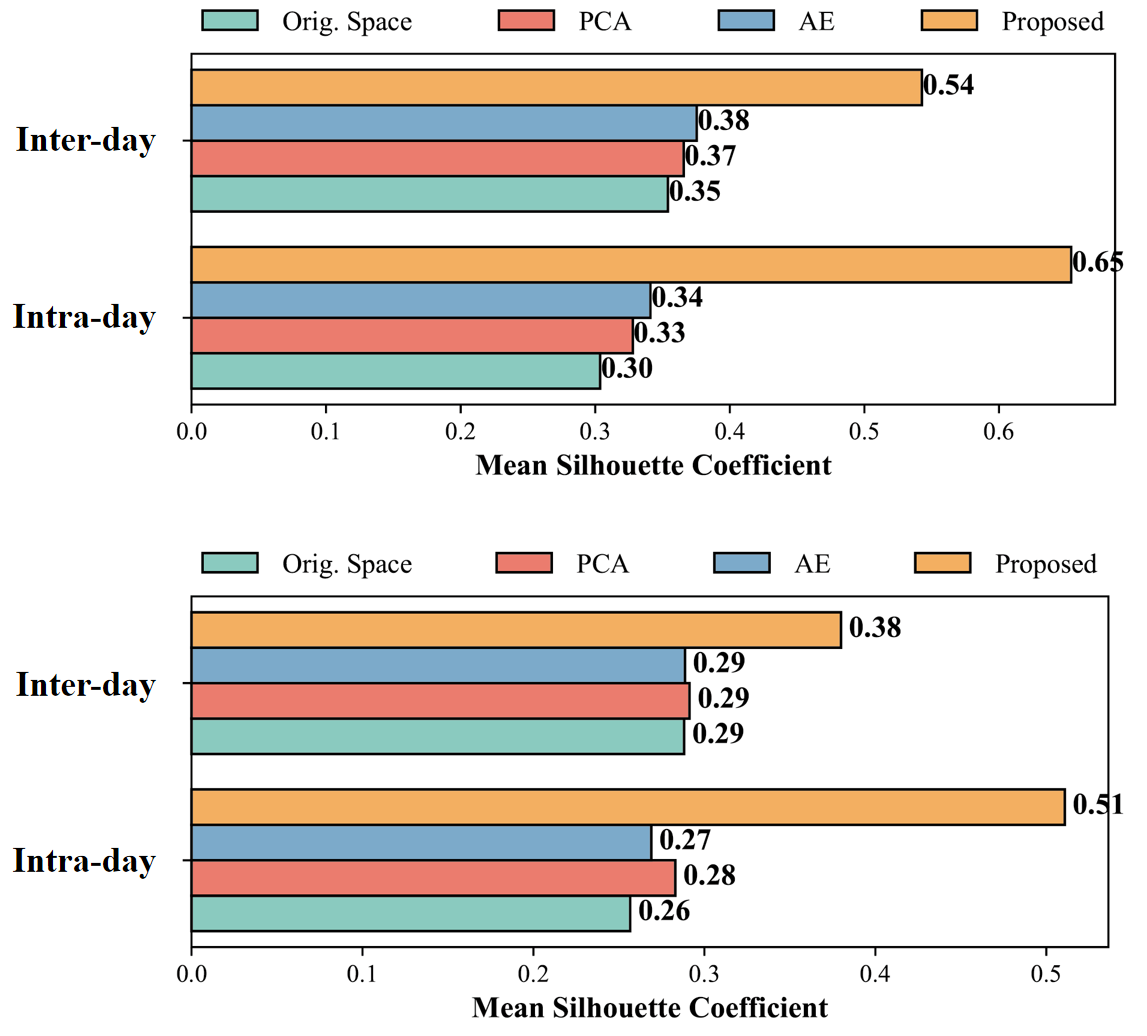}
    \caption{The SIL scores of the task- and subject-specific latent representations in intra-day and inter-day validation.}
    \label{fig:SIL_score}
\end{figure}

Figure \ref{fig:disentangle_tsne} illustrates the latent representations visualized by t-distributed Stochastic Neighbor Embedding (t-SNE). In both the pattern and identity latent spaces, features without disentanglement are chaotically distributed. In contrast, the task-specific and subject-specific components decomposed by our method are well-clustered in latent spaces under both intra-day and inter-day validation.  In addition, Figure \ref{fig:SIL_score} provides quantitative insights into the clustering structure. In both the gesture and identity spaces, the derived task- and subject-specific components consistently show the highest Silhouette Index (SIL) scores among all the approaches.

\subsection{RMS Maps of Original signals and Reconstructed Components}

\begin{figure*}[htbp]                        
    \centering
    \includegraphics[width=0.9\linewidth]{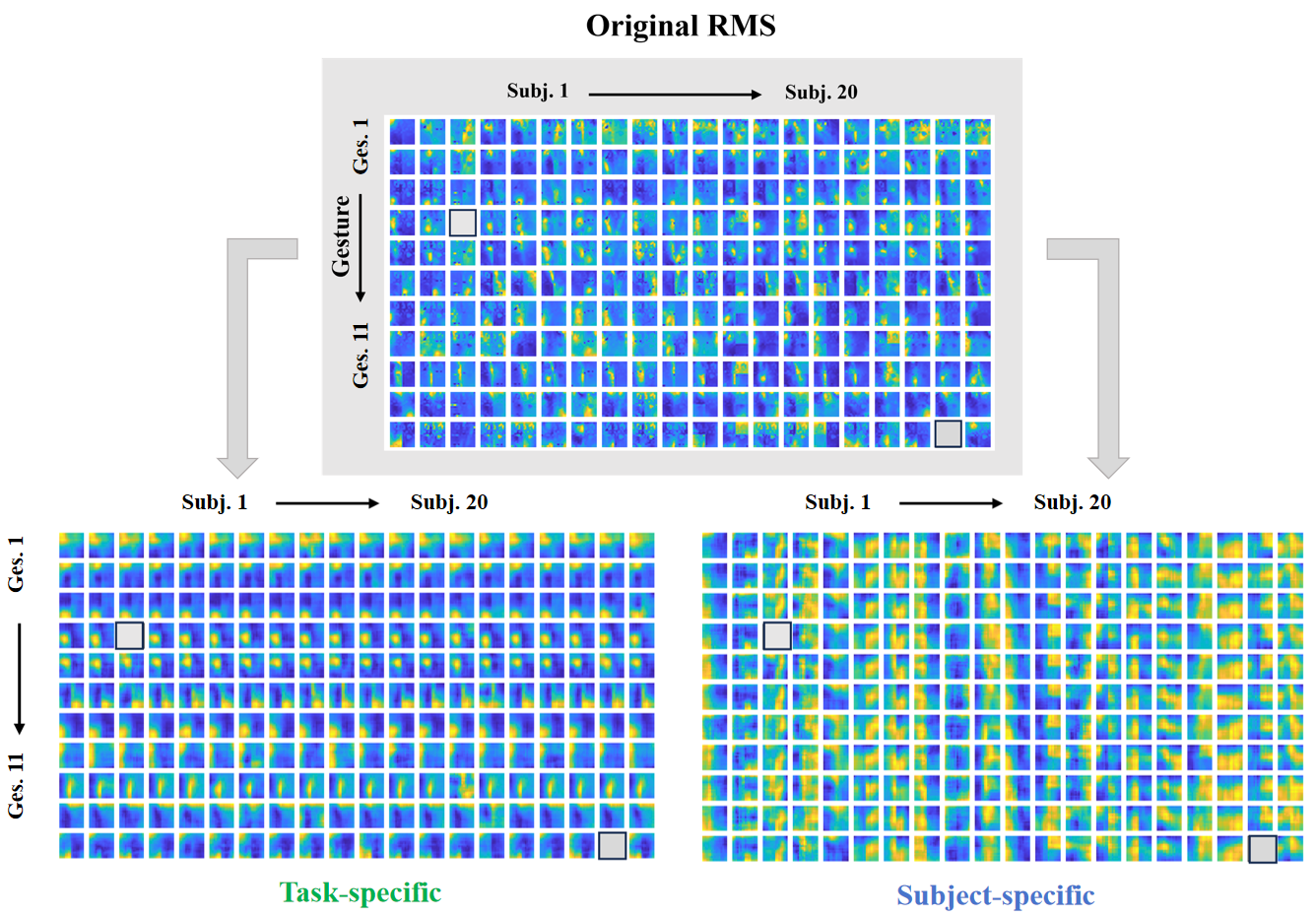}
    \caption{The visualization of the original RMS (top) and the RMS reconstructed by the proposed approach (bottom left denotes the task-specific reconstruction, while bottom right indicates the subject-specific reconstruction). The grids in the same \textcolor{black}{column} represent one subject, whereas the grids in the same \textcolor{black}{row} represent the same gesture. Ges. and Subj. denote Gesture and Subject respectively. For enhanced readability, each RMS map was interpolated to a scale of 5 using cubic interpolation. The gray square represents data from corresponding gestures and subjects was not available. }
    \label{fig:disentangle_RMS}
\end{figure*}

The RMS maps in Figure \ref{fig:disentangle_RMS} illustrate the localized spatial activation patterns associated with each 1-DoF hand gesture and each subject. \textcolor{black}{The reconstructed sEMG signal containing only task-specific or subject-specific component is generated by replacing one component with an all-zero matrix of the same shape and then inputting it into the decoder along with another component.} In the original RMS (top in Figure \ref{fig:disentangle_RMS}), distinct patterns can be observed between both the same subject for different gestures and the same gesture for different subjects. This variability is expected, as EMG amplitudes are influenced not only by muscle activation patterns but also by individual physiological differences.

Distinguished from the original RMS, in the reconstructed RMS array of task-specific components (bottom left in Figure \ref{fig:disentangle_RMS}), the heatmaps in the same column exhibit different patterns, whereas those in the same row are very similar. This indicates that the RMS of the task-specific components preserves the differences between gestures while showing high consistency for the same gesture across different subjects. In contrast, the reconstructed RMS of subject-specific components (bottom right in Figure \ref{fig:disentangle_RMS}) shows similar distributions within the same column but significant differences across the same row. Therefore, subject-specific components, as their name suggests, represent the varied characteristics of different individuals, preventing them from the influence of different gestures.

\begin{figure*}[htbp]
    \centering 
    \includegraphics[width=0.9\linewidth]{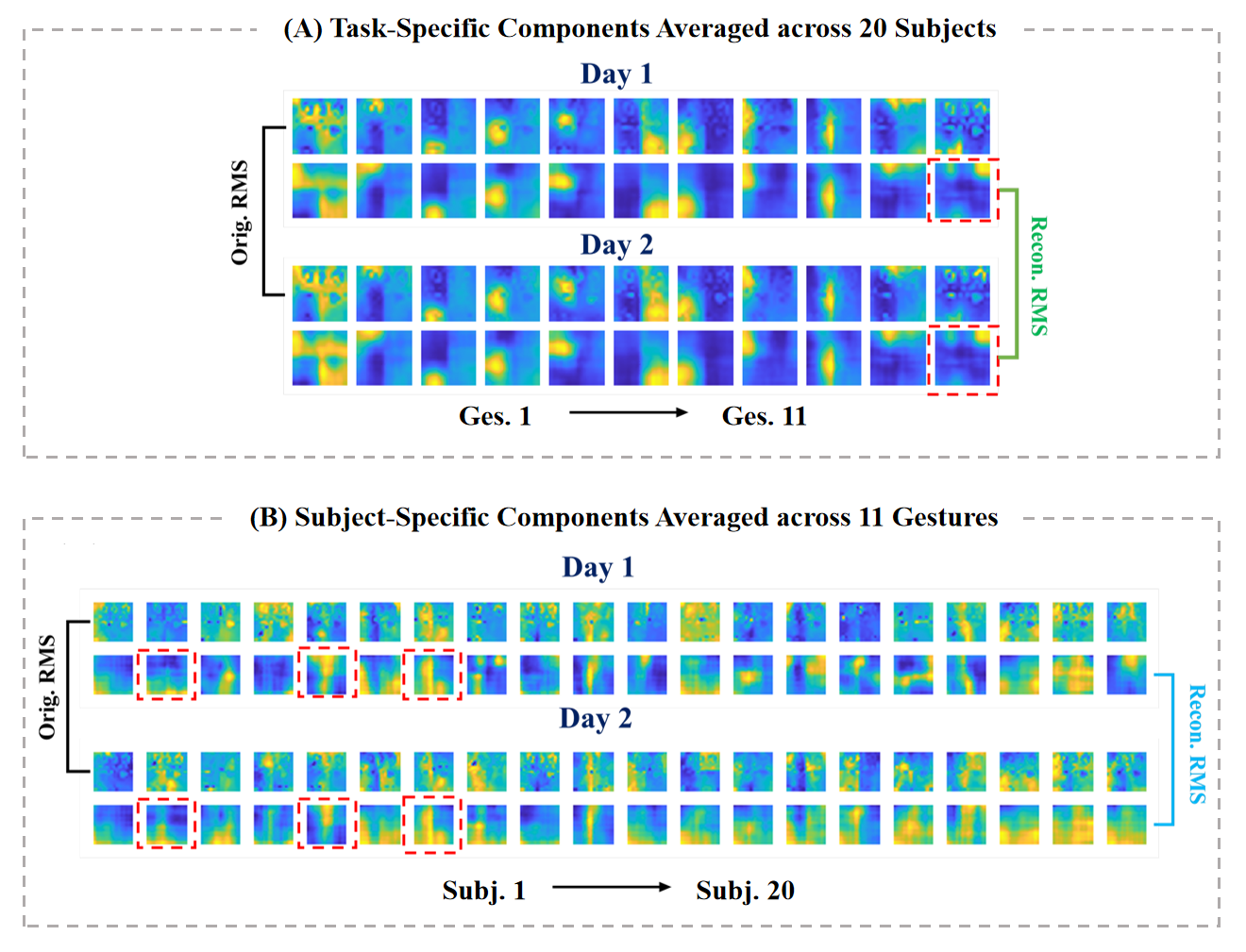}
    \caption{The averaged 'task' RMS and 'identity' RMS. Ges. and Subj. denote Gesture and Subject respectively. (A) the overlap of RMS maps for the same gesture across all subjects, and (B) the overlap of RMS maps for the same identities across the performed different gestures. The RMS maps were acquired on two separate days. The encircled grids demonstrate clear consistency of the reconstructed RMS between different days, which is beneficial for inter-day gesture recognition or classification}
    \label{fig:compare_interday_RMS}
\end{figure*}

Figure \ref{fig:compare_interday_RMS} (A) presents the average RMS map (across different subjects) of the task-specific components and the original signals from different days. It can be observed that the RMS maps of the task-specific components exhibit higher similarity between the two days compared to the original signals. In contrast, Figure \ref{fig:compare_interday_RMS} (B) presents the average RMS map (across different gestures) derived from the subject-specific components and the original RMS obtained on different days. In most cases, particularly those highlighted in red, there is considerable consistency observed between the RMS maps of subject-specific components across different days, which is almost absent between those of the original signals. This indicates that our method is capable of capturing the inherent properties of motor control.

\subsection{Quantification of the Centroid Distribution in RMS Maps}

\begin{figure*}[htbp]
    \centering
    \includegraphics[width=0.8\linewidth]{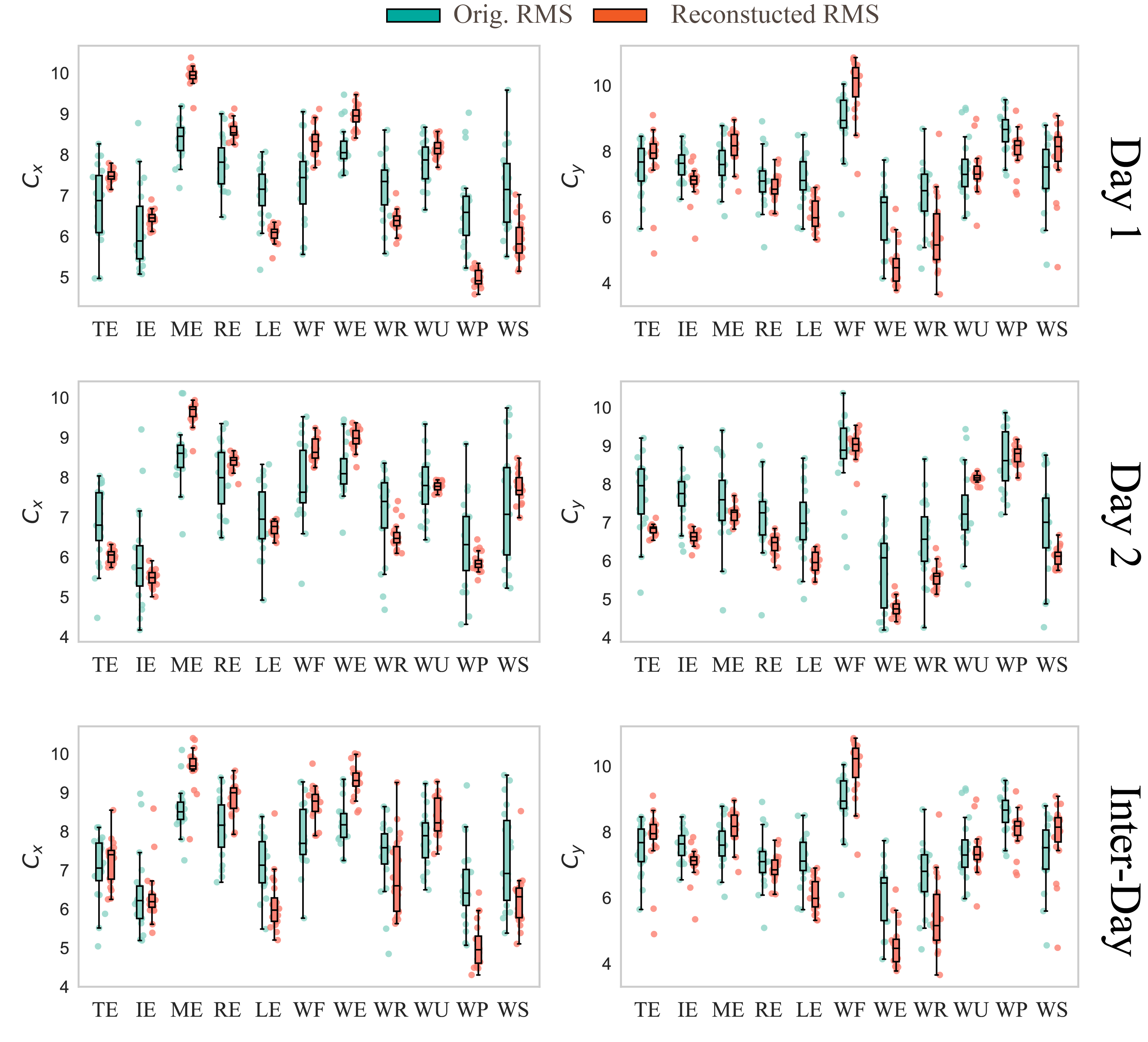}
    \caption{The comparison between the centroid distribution of each gesture across all subjects, as depicted by the original RMS and the RMS reconstructed by the disentangled task-specific component.}
    \label{fig:centroid_pattern}
\end{figure*}

\begin{figure*}[htbp]
    \centering
    \includegraphics[width=0.9\linewidth]{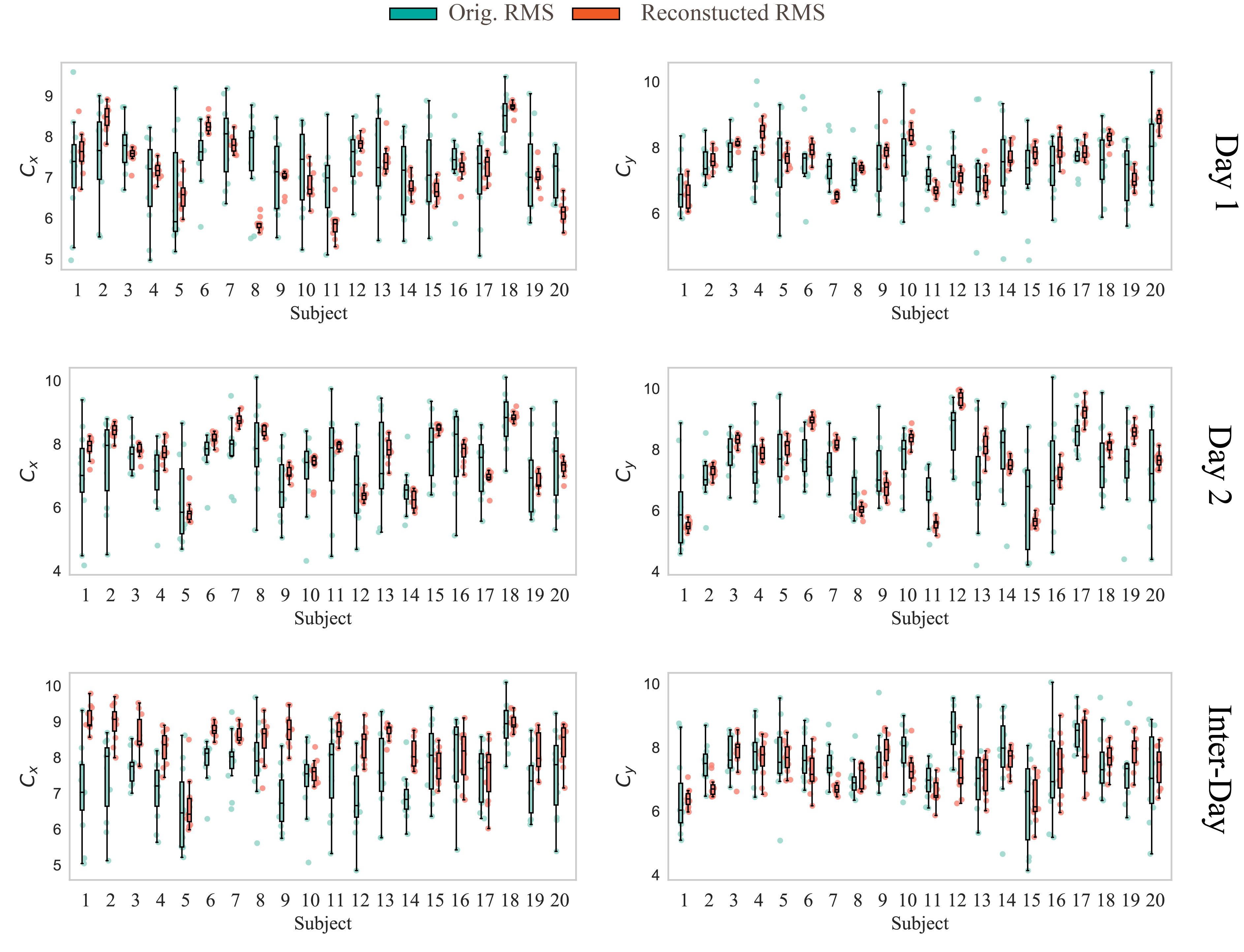}
    \caption{The comparison between the centroid distribution of each subject across all gestures, as depicted by the original RMS and the RMS reconstructed by the disentangled subject-specific component}
    \label{fig:centroid_indentity}
\end{figure*}

The centroid distribution of the RMS map for each gesture was calculated, measured with the distance from the centroid to the top-left corner of the RMS map. Figure \ref{fig:centroid_pattern} illustrates the mean and variance of the centroid distance across all subjects for each gesture. Compared to the original data, the disentangled task-specific components show smaller variance in centroid distance for each gesture. For mean values, differences exist in centroid distances across different gestures, and these differences are more pronounced for the task-specific components than for the original signals, especially in the intra-day scenario.

In contrast, Figure \ref{fig:centroid_indentity} illustrates the mean and variance of the centroid distance across all gestures for each individual. Similar to the task-specific components, the subject-specific components also exhibit greater variation in the mean distances compared to the original signals, while showing smaller variance for each individual.

\section{Discussion}

This study proposed an encoder-decoder-based disentanglement framework, extracting task-specific components and subject-specific components entangled in sEMG signals generated during hand movements. Furthermore, the neurophysiological characteristics of these two components are analyzed through visualization and quantification. The DoFs of the hand include extension and flexion of the five fingers, as well as wrist extension/flexion, radial/ulnar deviation, and pronation/supination. To avoid the complexity of multiple muscle activations associated with multi-DoF gestures, this study selected 11 single-DoF gestures from these DoFs, enabling a more fundamental analysis of the relationship between the extracted components and neuromuscular activities.

The proposed model in this study can be regarded as a useful sEMG feature extractor. The superiority of this extractor has been demonstrated through the classification performance validation and the physiological interpretability analysis of the extracted features. For classification accuracy, the use of disentangled task-specific and subject-specific features respectively achieved higher gesture classification and user identification accuracy compared to similar algorithms. \textcolor{black}{In a preliminary study, we selected KNN, Support Vector Machines (SVM) and Random Forest (RF) as typical machine learning methods, and Artificial Neural Network (ANN), Convolutional Neural Networks (CNN) and Long Short-Term Memory Networks (LSTM) as representative lightweight neural network for validation. The result show that all these classifiers, except LSTM, achieved equivalent recognition accuracies in the same task (both gesture recognition and user identification). The similar accuracy further indicates that the disentanglement model has already distinguished between different gestures or individuals at the feature extraction level, rather than relying on the performance of the classifier. LSTM performed worse than other classifiers probably because the temporal information has already been encoded into low-dimensional features by the encoders during disentanglement, making LSTM sensitive to temporal correlations.} Therefore, to highlight that the model extracts more generalized sEMG features instead of being optimized specifically for the classifier, this study deliberately adopts a simple linear classifier, KNN, for performance evaluation. As a complement to many studies focusing on improving the classification accuracy, the proposed feature extraction method can be integrated with any complex classifier in subsequent applications, further enhancing their recognition performance without requiring additional model calibration. \textcolor{black}{ Despite using only the simplest classifier, this study achieved a significant improvement in gesture recognition and identity authentication accuracy in inter-day scenarios, further validating the correctness of the disentanglement model. The results obtained are comparable to those from previous studies that used exhaustive feature selection \cite{jiang2022optimization} or advanced deep learning algorithms \cite{li2025deep} (such as transfer learning) on the same database, achieving over 90\% accuracy in inter-day gesture recognition task and nearly 70\% accuracy in identity authentication task \cite{jiang2021enhancing}. These accuracy levels are also consistent with results obtained from other inter-day validation databases. This demonstrates that the disentanglement model is capable of distinguishing between different gesture classes or individuals at the feature level. Moreover, the proposed algorithm is not incompatible with the use of more complex classification models. Future researchers can integrate their advanced models with the disentanglement model to further improve accuracy, and investigate the online performance of the model.}

% task-specific
In this study, task-specific components were compared to the neuromuscular activation patterns observed across a broad range of subjects, using the average muscle activation patterns of participants (original RMS) as a reference. In comparison with the original RMS, the activation patterns in the RMS maps showed more localized distributions, and they reflect the physiological and anatomical principles of the gestures performed. For example, the activation pattern for thumb finger extension is primarily observed in the most radial and distal regions, covering areas associated with the extensor pollicis brevis and extensor pollicis longus, which are responsible for thumb extension. Furthermore, the extensions of the index, middle, and ring fingers are distinctly observable along the distal-proximal direction. Specifically, the activation for index finger extension is localized on the distal end of the extensor digitorum communis, while the middle finger extension appears most proximally on the extensor digitorum communis. The activation for the ring and pinky fingers is situated between these two, showcasing the active compartments of the muscle during different finger movements. Furthermore, the box plot (Figure \ref{fig:centroid_pattern}) indicates that the activation regions extracted by the proposed algorithm are more focused, with greater distinctions in centroid positions. This enhanced variety of activation regions for different gestures makes great contributions to the accuracy of gesture recognition.

% subject-specific
For subject-specific components, the activation patterns are challenging to validate due to variations in individual force exertion habits. From the visualization results, distinct activation patterns exhibit across different individuals, whereas similar ones appear for the same individual (Figure \ref{fig:disentangle_RMS}). This outcome provides possible evidence for the reliability of the extracted subject-specific components. Regarding classification accuracy, user identification accuracy shows a notable decrease in inter-day validation compared to intra-day scenarios. Regarding the RMS maps, Figure \ref{fig:compare_interday_RMS} demonstrates that in inter-day conditions, the similarity of subject-specific components for the same individual is lower than that of task-specific components for the same gesture. This discrepancy may \textcolor{black}{result in} the more significant decrease in user identification accuracy than gesture classification accuracy in inter-day scenarios. The results suggest that electrode placement, data acquisition environment, and inter-day neuromuscular states significantly influence the representation of subject-specific components in sEMG signals, thus impacting inter-day user identification accuracy. \textcolor{black}{Additionally, previous studies\cite{li2021transfer} have shown that age can lead to significant changes in EMG features. Due to their strong association with individual characteristics, such changes are included in the subject-specific components during disentanglement. In the short term, age-related variability in muscle groups can serve as an individual feature that benefits user identification. In the long term, although such variability may cause the subject-specific components to perform worse across days, their isolation from the task-specific components enhances the generalization ability of the task-specific components across days and different age groups in gesture recognition tasks. As for inter-day user-identification tasks, the accuracy in this study is comparable to prior studies\cite{jiang2022measuring} on inter-day user identification with gestures, which often apply optimal feature selection or complex classifier models.} In contrast, this study employs an automated feature extraction algorithm and a simple KNN classifier. For real-world applications requiring higher inter-day user identification accuracy, combining multiple gestures could be an effective solution. Previous research \cite{jiang2021enhancing} has shown that combining four or more gestures can increase the identification accuracy to nearly 100\% when the accuracy with a single gesture is between 60-70\%, making sEMG-based identity recognition feasible for practical scenarios.

\textcolor{black}{The issue of electrode placement after replacement is a critical challenge in EMG-based human-machine interaction. The accuracy of gesture recognition inevitably decreases when the electrode placement changes. However, in practical use, the need for prolonged calibration each time due to the electrode replacement is not acceptable. Therefore, it is essential to account for the electrode displacement issue. Accordingly, this study presents results for the inter-day verification protocols. To minimize the impact of this factor, data augmentation during model training was applied to simulate electrode displacement (translation and rotation). The data augmentation method is consistent with our previous study\cite{jiang2022optimization}. However, the electrode placement position still inevitably affects the performance of disentanglement model. Nevertheless, both the inter-day gesture recognition accuracy and the inter-day identity authentication results show that the disentanglement model proposed in this study is comparable to the latest research using the same dataset\cite{meng2022user, jiang2022measuring}, which demonstrates the effectiveness of the model in practical applications. Future work can investigate advanced classification algorithms to further improve both the gesture recognition accuracy and identity authentication accuracy.}

Deep learning algorithms have been criticized in the biomedical field for their lack of interpretability. Building on the prior disentanglement model proposed by the research group\cite{fan2024surface}, this study further conducted a neurophysiological analysis of the disentangled features extracted by the deep model. \textcolor{black}{However, it should be acknowledged that this physiological validation is relatively superficial. Future studies can incorporate imaging or anatomical methods to further verify the physiological interpretability of the proposed algorithm.} sEMG signals are physiological electrical signals generated by human movements but are often influenced by intrinsic factors such as individual characteristics and extrinsic factors like environmental conditions. Currently, most work in this field focuses on improving motion recognition accuracy, which primarily addresses pattern-related components. The disentanglement model proposed in this study may open new avenues for future research by separating task-specific components from EMG signals to isolate subject-specific components. These subject-specific components may reveal additional information, including personal data for identification, pathological information for patients, and insights into individual mental or physical states. Such information could enable targeted applications of subject-specific components in specific scenarios, thereby enhancing the practical value of EMG signals in real-world applications. 

 \textcolor{black}{\section{Conclusion}
This study proposed a novel encoder-decoder-based disentanglement framework to extract task-specific and subject-specific components from sEMG signals generated during hand movements. The results demonstrated that the task-specific components retained consistent neuromuscular activation patterns across individuals, leading to significantly improved gesture classification accuracy, and the subject-specific components captured individual characteristics that enhanced identity recognition performance. The visualization and quantification of the disentangled components revealed distinct neuromuscular activation patterns that align with physiological and anatomical principles, further enhancing the interpretability of the model. Importantly, the proposed framework demonstrated its potential to improve classification accuracy without relying on complex classifiers, allowing future integration with advanced models to further enhance performance for online real-world applications.}

\section*{Declaration of competing interest}
The authors declare that they have no known competing financial interests or personal relationships that could have appeared to influence the work reported in this paper.

\section*{Declaration of generative AI and AI-assisted technologies in the writing process}
During the preparation of this work the authors used ChatGPT in order to improve language and readability. After using this tool/service, the authors reviewed and edited the content as needed and take full responsibility for the content of the publication.

% \clearpage
% \newpage
\bibliographystyle{elsarticle-num}
\bibliography{refs}

\end{document}